\documentclass[twocolumn,showpacs,superscriptaddress,floatfix,aps,prl]{revtex4}

\usepackage{amsmath}
\usepackage{graphicx}

\begin{document}

\author{P. Sebbah}
\email[Contact: ]{sebbah@unice.fr} \affiliation{Department of
Physics, Queens College of the City University of New York,
Flushing, New York 11367} \affiliation{Laboratoire de Physique de
la Mati{\`{e}}re Condens{\'{e}}e, Universit{\'{e}} de Nice -
Sophia Antipolis,
\\Parc Valrose, 06108, Nice Cedex 02, France}

\author{B. Hu}
\affiliation{Department of Physics, Queens College of the City
University of New York, Flushing, New York 11367}

\author{J. M. Klosner}
\affiliation{Department of Physics, Queens College of the City
University of New York, Flushing, New York 11367}

\author{A. Z. Genack}
\affiliation{Department of Physics, Queens College of the City
University of New York, Flushing, New York 11367}

\title{Extended quasimodes within nominally localized random waveguides}
\date{\today}

\begin{abstract}
We have measured the spatial and spectral dependence of the
microwave field inside an open absorbing waveguide filled with
randomly juxtaposed dielectric slabs in the spectral region in
which the average level spacing exceeds the typical level width.
Whenever lines overlap in the spectrum, the field exhibits
multiple peaks within the sample. Only then is substantial energy
found beyond the first half of the sample. When the spectrum
throughout the sample is decomposed into a sum of Lorentzian lines
plus a broad background, their central frequencies and widths are
found to be essentially independent of position. Thus, this
decomposition provides the electromagnetic quasimodes underlying
the extended field in nominally localized samples. When the
quasimodes overlap spectrally, they exhibit multiple peaks in
space.

\end{abstract}


\maketitle

The nature of wave propagation in disordered samples reflects the
spatial extent of the wave within the medium.
\cite{Anderson,Thouless,gangof4} In part because of the
inaccessibility of the interior of multiply-scattering samples,
the problem of transport in the presence of disorder has been
treated as a scattering problem with the transition from extended
to localized waves charted in terms of characteristics of
conductance and transmission. When gain, loss, and dephasing are
absent, the nature of transport can be characterized by the degree
of level overlap, $\delta$, \cite{Thouless,gangof4} which is the
ratio of average width and spacing of states of an open random
medium, $\delta\equiv\delta\nu/\Delta\nu$. Here $\delta\nu$ may be
identified with the spectral width of the field correlation
function \cite{Genack} and $\Delta\nu$ with the inverse of the
density of states of the sample. When $\delta > 1$, resonances of
the sample overlap spectrally, the wave spreads throughout the
sample and transport is diffusive. \cite{Thouless} In contrast,
when $\delta < 1$, coupling of the wave in different portions of
the sample is impeded. Azbel showed that,when $\delta < 1$,
transmission may occur via resonant coupling to exponentially
peaked localized modes with spectrally isolated Lorentzian lines
with transmission approaching unity when the wave is localized
near the center of the sample. \cite{Azbel,Freilikher} Given the
sharp divide postulated between extended and localized waves, the
nature of propagation when modes occasionally overlap in samples
for which $\delta < 1$ is of particular interest.

Mott argued that interactions between closely clustered levels in
a range of energy in which $\delta < 1$ would be associated with
two or more centers of localization within the sample within the
sample. \cite{Mott} Pendry showed that, in this case, occasionally
overlapping of electronic modes play an outsized role in transport
since electrons may then flow through the sample via regions of
high intensity which are strung together like beads in a necklace.
\cite{PendryPhysC,PendryAdvPhys} Recent pulsed \cite{Wiersma} and
spectral \cite{Wiersma,Milner} measurements of optical
transmission in layered samples were consistent with the
excitation of multiple resonances associated with necklace states.
In related work, Lifshits had shown that the hybridization of
overlapping defect states outside the allowed vibrational bands of
the pure material led to the formation of a band of extended
states. \cite{Lifshits} Similar impurity bands can be created for
electrons in the forbidden gap of semiconductors, photons in
photonic band gaps, and polaritons in the polariton gap
\cite{Deych}.

The spatial distribution of Lorentzian localized modes has been
observed experimentally in a one dimensional loaded acoustic line
systems in which a single localized mode is excited and the
spectral line is Lorentzian \cite{Maynard} but the spatial
distribution of the field has not been observed for necklace
states when many lines hybridize. The question then arises as to
whether the field distribution in an open dissipative system may
be expressed as a superposition of decaying quasimodes. If so, the
quasimodes underlying the necklace states could be observed
allowing for an exploration of their spatial and spectral
characteristics such as a comparison of the widths of these modes
to the sum of the leakage and dissipation rates, and a
consideration of their completeness and orthogonality.
\cite{Petermann,Siegman, Ching, Dutra,Schomerus} Related issues
are relevant to quasimodes of decaying nuclei, atoms and
molecules, electromagnetic waves in microspheres or chaotic
cavities, and gravity waves produced by mater captured by black
holes. \cite{Ching}

In this Letter, we explore the nature of propagation in an open
dissipative random one-dimensional dielectric medium embedded in a
rectangular waveguide. Measurements of the microwave field
spectrum made at closely spaced points along the entire length of
the waveguide over a spectral range in which $\delta<1$, reveals
the spatial distribution of the field. In all samples, we find
spectrally isolated Lorentzian lines associated with exponentially
localized waves. Whenever peaks in the field overlap in space,
they also overlap spatially. Only then does the field penetrates
substantially beyond the first half of the sample. When the field
spectra at each points within the sample are decomposed into a sum
of Lorentzian lines and a slowly varying background, the central
frequencies and widths of the lines are found to be independent of
position within the sample.

\begin{figure}
\centering
\includegraphics[width=8cm,angle=0]{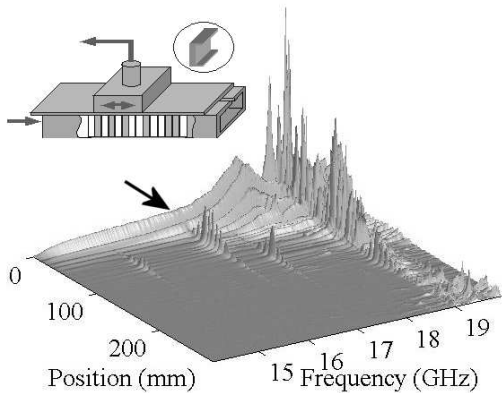}
\caption{Spectra of the field amplitude at each point along a
typical random sample normalized to the amplitude of the incident
field. The arrow points to the input direction. Inset: Description
of the experimental set up and schematic of the ceramic structure
element.} \label{fig1}
\end{figure}

Field spectra are taken along the length of a slotted W-42
microwave waveguide using a vector network analyzer. The field is
weakly coupled to a cable without a protruding antenna inside a
copper enclosure. A 2 mm-diameter hole in the enclosure is pressed
against the slot, which is otherwise covered by copper bars
attached to both sides of the enclosure (inset Fig.~\ref{fig1}).
The entire detector assembly is translated in 1 mm steps by a
stepping motor. Measurements are made in one hundred random sample
realizations each composed of randomly positioned dielectric
elements. Ceramic blocks are milled to form a binary element of
length $a = 7.74 \pm 0.04$ mm. The first half of the block is
solid and the second half comprises two projecting thin walls on
either side of the air space, as shown in the inset of
Fig.~\ref{fig1}. The orientation of the solid element towards or
away from the front of the waveguide is randomly selected. This
structure introduces states into the band gap of the corresponding
periodic structure \cite{John} close to the band edges. In order
to produce states in the middle of the band gap, ceramic slabs of
thickness $a/2$, corresponding to the solid half of the binary
elements, or Styrofoam slabs, with refractive index close to
unity, are inserted randomly. The samples are composed of 31
binary elements, 5 single ceramic elements and 5 Styrofoam
elements with a total length of 28.8 cm.

\begin{figure}
\centering
\includegraphics[width=8cm,angle=0]{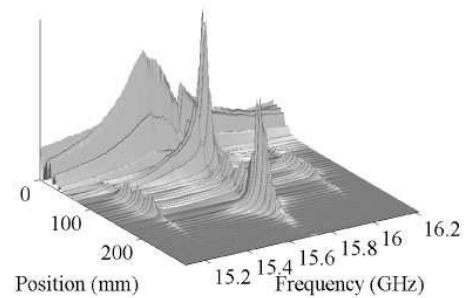}
\includegraphics[width=8cm,angle=0]{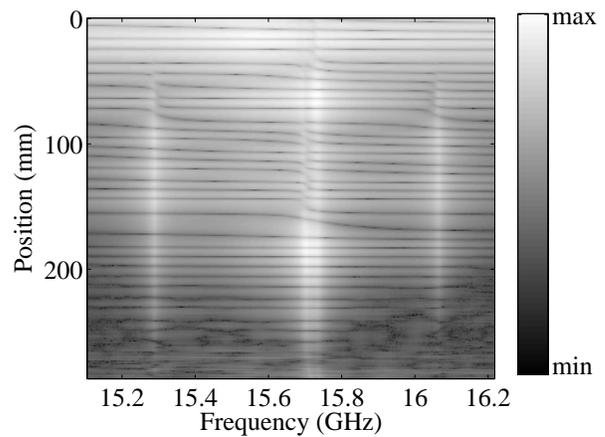}
\caption{(a) Spectra of the field amplitude at each point along a
random sample with spectrally overlapping peaks normalized to the
amplitude of the incident field. (b) The top view of Fig.
\ref{fig2}a in logarithmic presentation.} \label{fig2}
\end{figure}

Spectra of the field amplitude at equally spaced points along a
typical random sample configuration normalized to the amplitude of
the incident field amplitude are shown in Fig.~\ref{fig1}. A few
isolated exponentially peaked localized modes with Lorentzian
linewidths are seen below 18.7 GHz. Between 18.7 GHz and 19.92
GHz, lines generally overlap so that $\delta > 1$ and the wave is
extended. The field amplitude within the sample over a narrower
frequency range in a different random configuration in which
spectrally overlapping peaks are observed within the band gap, is
shown in two projections in Fig.~\ref{fig2}.
In order to exhibit the progression of the phase within the
sample, a top view of the logarithm of the amplitude of the data
plotted in Fig.~\ref{fig2}a is presented in Fig.~\ref{fig2}b. In
the periodic binary structure, the same projection as in
Fig.~\ref{fig2}b, shows parallel ridges corresponding to maxima of
the field amplitude separated by $a$. Thus half the wavelength
equals $a$ throughout the band gap, or $\lambda$ = 2$a$. When the
frequency is tuned through a Lorentzian line, which corresponds to
a localized state (e.g. $\nu=15.3$ GHz), the phase through the
sample increases by $\pi$ rad and an additional peak in the
amplitude variation across the sample is introduced. In the
frequency interval between 15.6 and 15.8 GHz where multiple peaks
are observed in the field distribution, the number of ridges
increases by 3.
\begin{figure}
\centering
\includegraphics[width=6.6cm,angle=-90]{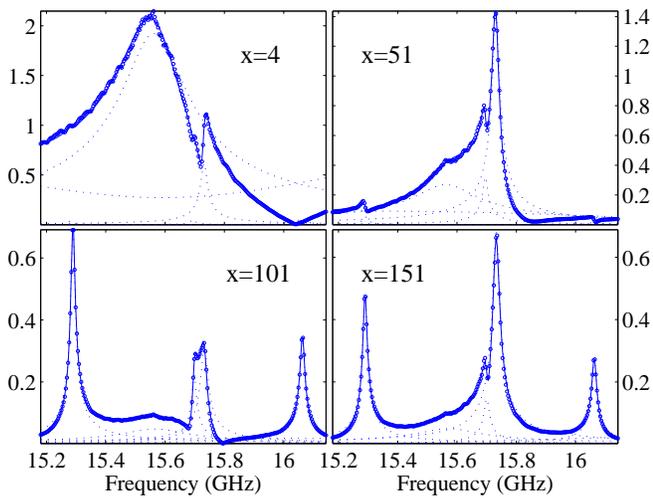}
\caption{Comparison of the measured field magnitude (dots) to
Eq.~\ref{Eq1} (full line) at four different locations. The
Lorentzian lines and the polynomial in Eq.~\ref{Eq1} are
represented in dotted lines.} \label{figMerit}
\end{figure}
This suggests that the radiation is tuned through the central
frequency of three successive resonances. This is tested by
fitting the field spectrum to a sum $N$ of Lorentzian lines, as
follows,
\begin{equation}
\label{Eq1} E(\nu ,x) = \sum\limits_{n = 1}^N {\frac{A_n
(x)}{{\Gamma _n (x) + i(\nu  - \nu _n (x))}} + \sum\limits_{m =
0}^2 {C_m (x)(\nu  - \nu _0 )^m }}
\end{equation}
where $A_n (x)$ and $C_n (x)$ are complex coefficients. The slowly
varying polynomial of the second degree centered at $\nu_0$
represents the sum of the evanescent wave and the tail of the
response of distant lines. Here, $\nu_0=15.66$ GHz, which is the
center of the frequency interval considered. We choose $N=5$ to
include the two ``satellite'' lines at 15.3 GHz and 16.06 GHz. An
iterative double least-squares fit procedure is applied
independently at each position $x$ within the sample as follows:
first guesses for the central frequencies, $\nu_n$, and linewidths
$\Gamma_n$ are used to fit the spectrum measured at each position
$x$ to Eq.~\ref{Eq1} with the amplitude coefficients $A_n(x)$, and
$C_n(x)$ as fitting parameters. These values are used in a second
step in which, only the $\nu_n(x)$, and $\Gamma_n(x)$ are fitting
parameters within a bounded spectral range. This double fitting
procedure can be repeated to improve the fit. The quality of the
fit can be seen in Fig.~\ref{figMerit}. Indeed, the $\chi^2$
normalized by the product of the integrated spectrum and degrees
of freedom (the number of point minus the number of free
parameters), remains below $2\times10^{-3}$ over $90\%$ of the
sample length. The noise is higher near the sample output where
the signal is generally close to the noise level. The central
frequencies, $\nu_n(x)$, and linewidths $\Gamma_n(x)$, found in
the fit are shown in Fig.~\ref{fig3}. A plot of the complex square
of each term in Eq.~\ref{Eq1} is shown in Fig.~\ref{fig4}.
Fluctuations in $\nu_n(x)$ and $\Gamma_n(x)$ are large only at
positions for which the peak magnitude of the terms for the $n$th
mode are low, as can be seen by comparing Fig.~\ref{fig3} and
~\ref{fig4}. In the domain in which fluctuations in the central
frequencies and linewidths for a particular quasimode are low,
these quantities are virtually independent of position and the
field amplitude is given to good accuracy by substituting the
average values of $\nu_n(x)= \nu_n$ and $\Gamma_n(x)= \Gamma_n$ in
Eq.~\ref{Eq1}. The drift is greatest in mode 2, where the
variation in $\nu_2(x)$ is less than 20\% of $\Gamma_2$.

With $\nu_n$ and $\Gamma_n$ specified, each of the Lorentzian
terms in Eq.~\ref{Eq1} corresponds to a quasimode. The Fourier
transform of each term gives the response to an incident pulse in
which the temporal and spatial variation factorize,
$f_n(x,t)=f_n(x)\text{exp}(-i(2\pi\nu_n-i\Gamma_n)t)$
corresponding to a sum of exponentially decaying quasimodes. The
independent decay of the quasimodes indicates that they are
orthogonal.
\begin{figure}
\centering
\includegraphics[width=9.5cm,angle=0]{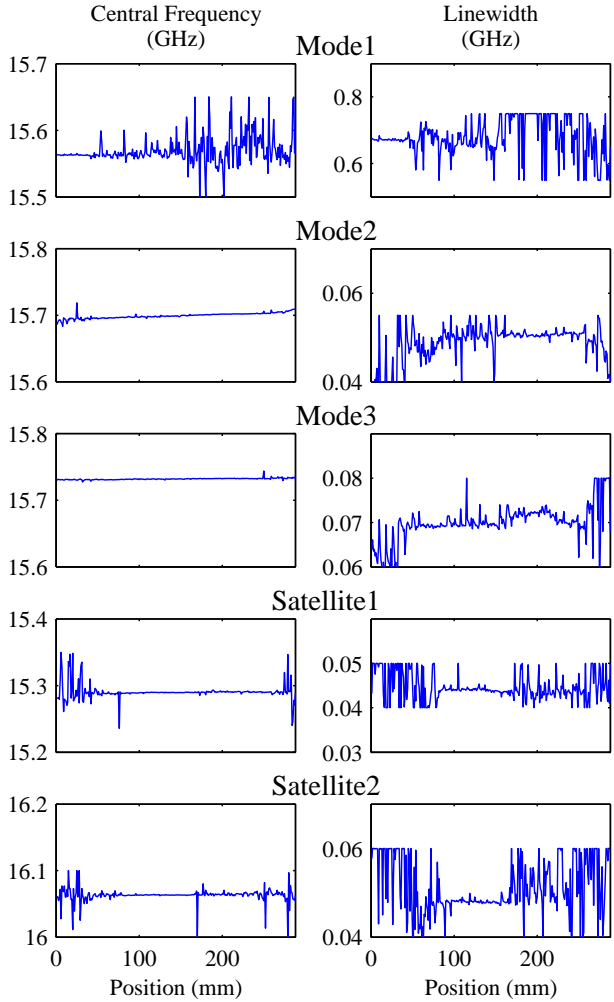}
\caption{Spatial dependence of central frequency and linewidth for
the five modes shown in Fig.~\ref{fig4}. At positions at which the
mode amplitude is greater than the noise, the central frequency
and linewidth are sharply defined.} \label{fig3}
\end{figure}
\begin{figure}
\centering
\includegraphics[width=8cm,angle=0]{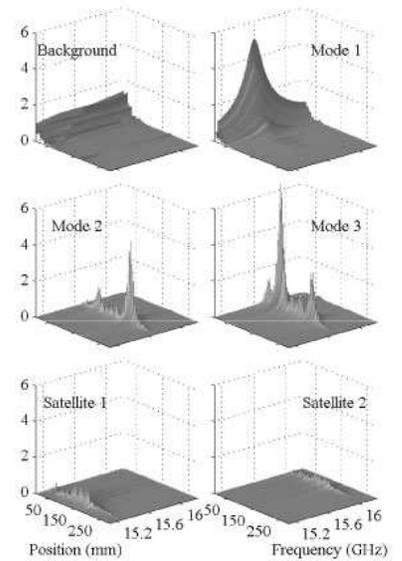}
\caption{Field magnitude for five quasimodes and slowly varying
polynomial term in Eq.~\ref{Eq1}.} \label{fig4}
\end{figure}
For the three spectrally overlapping modes, ``mode 1''
($\nu_1\simeq 15.56$ GHz, $\Gamma_1\simeq 0.67$ GHz) is broad as a
result of its closeness to the input. ``Mode 2'' ($\nu_2\simeq
15.70$ GHz, $\Gamma_2\simeq 0.051$ GHz) and ``mode 3''
($\nu_3\simeq 15.73$ GHz, $\Gamma_3\simeq 0.07$ GHz) are spatially
extended and multi-peaked in a spectral range in which quasimodes
are otherwise strongly localized. In fact, "satellite modes" are
not strictly localized due to the overlap with other modes. The
sharply defined modes appearing in the expansion of Eq.~\ref{Eq1}
in the regions in which the amplitude of specific modes is greater
than the noise in the measured field and the excellent fit to the
spectra shown in Fig.~\ref{figMerit} is consistent with these
modes representing a complete set even when they overlap.

Because of dissipation within and leakage from the random sample,
the system is not Hermitian, but it remains symmetric since
reciprocity is preserved. A non-Hermitian but symmetric
Hamiltonian has complex orthogonal eigenvalues when the
appropriate inner product is used. \cite{Ching,Weaver} If the
quasimodes found in the fit and shown in Fig.~\ref{fig4} were
orthogonal, the linewidths of quasimodes would equal the sum of
the absorption rate, $\Gamma_a$ and the leakage rates for specific
modes, $\Gamma_{ln}$, for the $n$th quasimode. The sample
dissipation rate is given by the ratio of the net flux into the
sample, which is the difference between the incident flux and the
sum of the reflected and transmitted fluxes, and the steady-state
electromagnetic energy in the sample, which is proportional to
$\int_0^L\epsilon(x)E^2dx$. Here $\epsilon(x)$ is the effective
relative permittivity determined by matching the measured
frequencies of the modes at the band edge of the periodic
structure with the results of one-dimensional model and including
waveguide dispersion. $\Gamma_{ln}$ is determined from the ratio
of flux away from the sample for a given quasimode to the energy
in the sample for this quasimode for steady-state excitation. The
leakage from the sample for a given quasimode is obtained by
decomposing the scattered wave in the empty waveguide before and
after the random sample into a sum of Lorentzian lines as given in
the first sum on the right hand side of Eq. (\ref{Eq1}) . The flux
is then proportional to the product of the square of the amplitude
of this field component and the group velocity in the waveguide.
Within experimental uncertainty of 25$\%$, we find that the mode
linewidths are equal to the sum of the absorption and leakage
rates. This is consistent with the conclusion that quasimodes
found are orthogonal.

In conclusion, we have shown that the field inside an open
dissipative random single-mode waveguide can be decomposed into a
complete set of quasimodes, even when the mode spacing is
comparable to the linewidths. Smaller mode spacing do not occur
for spatially overlapping mode because of mode repulsion. We find
a multi-peaked extended field distribution whenever quasimodes
overlap, and a single peaked field distribution in the rare cases
of spectrally isolated modes. Because absorption suppresses
long-lived single-peaked states more strongly than short-lived
multiply peaked states, wave penetration into the second half of
the sample and transmission through the sample are substantial
only when a number of closely spaced quasimodes are excited.
Though waves may extend through the sample when quasimodes overlap
in a spectral range in which $\delta <1$, their envelope still
falls exponentially near each peak representing a center of
localization and the nature of transport may still be
differentiated from diffusive regions for which $\delta >1$. The
demonstration that quasimodes are well-defined when the spacing
between their central frequencies is comparable to their
linewidth, suggests that a quasimode description may also be
appropriate for diffusing waves in samples with $\delta > 1$.

We thank H. Rose, Z. Ozimkowski and L. Ferrari for suggestions
regarding the construction of the waveguide assembly, and V.
Freilikher, J. Barthelemy, L. Deych, V.I. Kopp, O. Legrand, A.A
Lisyansky, F. Mortessagne, and R. Weaver for valuable discussions.
This research is sponsored by the National Science Foundation
(DMR0205186), by a grant from PSC-CUNY, by the CNRS (PICS
$\#2531$) and is supported by the Groupement de Recherches IMCODE.


\end{document}